\documentclass[reprint, amsmath, amssymb, superscriptaddress, floatfix, showpacs, showkeys, aps, prb]{revtex4-1}

\usepackage{graphicx}
\usepackage{float}
\usepackage{color}
\begin{document}
\preprint{}

\title{Paramagnetic to ferromagnetic phase transition in lightly Fe-doped Cr$_2$B}

\author{Leslie Schoop}
\email{lschoop@princeton.edu}
\affiliation{Department of Chemistry, Princeton University, Princeton New Jersey 08544, USA.}
\affiliation {Max-Planck-Institut f\"ur Chemische Physik fester Stoffe, 01187 Dresden, Germany.}
\author{Max Hirschberger}
\affiliation{Joseph Henry Laboratory, Department of Physics, Princeton University, Princeton New Jersey 08544, USA.}
\author{Jing Tao}
\affiliation{Department of Physics, Brookhaven National Laboratory, Upton NY 11973.}
\author{Claudia Felser}
\affiliation {Max-Planck-Institut f\"ur Chemische Physik fester Stoffe, 01187 Dresden, Germany.}
\author{N.P. Ong}
\affiliation{Joseph Henry Laboratory, Department of Physics, Princeton University, Princeton New Jersey 08544, USA.}
\author{R. J. Cava}
\affiliation{Department of Chemistry, Princeton University, Princeton New Jersey 08544, USA.}

\date{\today}

\begin{abstract}

Cr$_2$B displays temperature independent paramagnetism. We induce ferromagnetism by replacing less than $3\,\%$ of the Cr atoms by Fe. By the lowest Fe doping level made, Curie-Weiss behavior is observed; $\Theta_{CW}$ changes from $-20\,$K for $0.5\,\%$ Fe-doped Cr$_2$B to positive values of about 50~K by $5\,\%$ Fe doping. The ferromagnetic T$_C$ is 8 K for $2.5\,\%$ Fe doping and increases linearly to 46 K by $5\,\%$ doping; we infer that a quantum phase transition occurs near the $2.0\,\%$ Fe level. Magnetic fluctuations at the intermediate doping levels are reflected in the linear resistance and an anomalous heat capacity at low temperatures. Imaging and chemical analysis down to the atomic scale show that the Fe dopant is randomly distributed.

\end{abstract}


\maketitle

%%%%%%%%%%%%%%%%%%%%%%%%%%%%%%%%%%%%%%%%%%%%%%%%%%%%%%%%%%%%%%%%%%%%%%
\section{Introduction}
Materials that are close to a magnetic, electronic or structural instability remain the focus of many fundamental studies in condensed matter research. Phase transitions between two nearly degenerate magnetic ground states at low temperatures can be induced by tuning parameters such as pressure, doping, or the application of a magnetic field and often show quantum fluctuations that influence the behavior of thermodynamic and transport properties \cite{coleman2005quantum,keimer2011quantum}. Such quantum phase transitions (QPTs) for metallic magnets have been widely studied in materials with \textit{4f} electrons \cite{gegenwart2008quantum}. QPTs in \textit{3d} metals can also be expected \cite{Moriya1985,Lonzarich1985}, though comparatively few materials in this class are known. Among these are the ferromagnets ZrZn$_2$ \cite{uhlarz2004quantum}, Ni$_3$Al \cite{niklowitz2005spin}, MnSi \cite{Pfleiderer1997}, Pd$_{1-x}$Ni$_x$ \cite{nicklas1999non} and NbFe$_2$ \cite{brando2008logarithmic}. For antiferromagnetic QPTs, the most widely studied example is Cr$_{1-x}$V$_x$ \cite{yeh2002quantum}. Cr-containing materials are attractive candidates for observing such behavior as the metal itself is an itinerant antiferromagnet; the magnetic order can be suppressed with external pressure and a QPT is reached at 10 GPa \cite{jaramillo2010signatures}. 

Here we report a phase transition from paramagnetism to weak ferromagnetism in very lightly Fe-doped Cr$_2$B, (Cr$_{1-x}$Fe$_x$)$_2$B \cite{kayser1999re}. We find undoped Cr$_2$B to be paramagnetic down to 2 K, although its ground state is calculated to be antiferromagnetic \cite{zhou2009first}. The observed transition from a paramagnetic to a ferromagnetic state near $x = 0.02$ in this system at low temperatures displays some of the phenomenology associated with QPTs. 

\section{Experimental}

Polycrystalline samples of (Cr$_{1-x}$Fe$_{x}$)$_2$B were prepared up to \textit{x} = 0.05 from stoichiometric mixtures of the elements by arc melting under an Ar atmosphere and subsequent annealing. The buttons were wrapped in Ta foil and sealed in a quartz tube, where they were annealed at 1150 $^\circ$C for 72 h. (The Ni-doped samples for comparison were prepared the same way.) The purity of the samples was confirmed with powder X-ray diffraction on a Bruker D8 Focus x-ray diffractometer operating with Cu K$\alpha$ radiation and a graphite diffracted beam monochromator. In addition to the Cr$_2$B phase, a small amount (less than 5\%) of Cr metal was present in some of the samples. By similarly preparing samples of Fe-doped elemental Cr, we confirmed that their magnetization is negligible compared to that of the bulk doped Cr$_2$B, thus the magnetic characterization of the Fe doped Cr$_2$B phase is not affected. Heat capacity and magnetization measurements were performed with a Quantum Design Physical Property Measurement System (PPMS). The samples for resistivity measurements were polished platelets of approximate geometry $3\,$ mm $\times1.8\,$ mm $\times0.1\,$mm; gold pads were evaporated onto the platelets after argon etching, and contact to the pads was made using thin Au wire and silver paint. Temperature-dependent resistivities were measured in the PPMS or in an in-house cryostat. High resolution transmission electron microscopy (HRTEM) and high angle annular dark-field scattering transmission microscopy (HAADF-STEM) measurements were performed on a JEOL-ARM200F TEM with the accelerating voltage of 200 kV. The microscope has double Cs correctors to achieve high real-space resolution in both the HRTEM and HAADF-STEM modes with the energy resolution about 0.4 eV in the electron energy loss spectroscopic (EELS) results.

\section{Results and Discussion}

\subsection{Magnetism}

\begin{figure}[ht]
  \centering
  \includegraphics[width=\columnwidth]{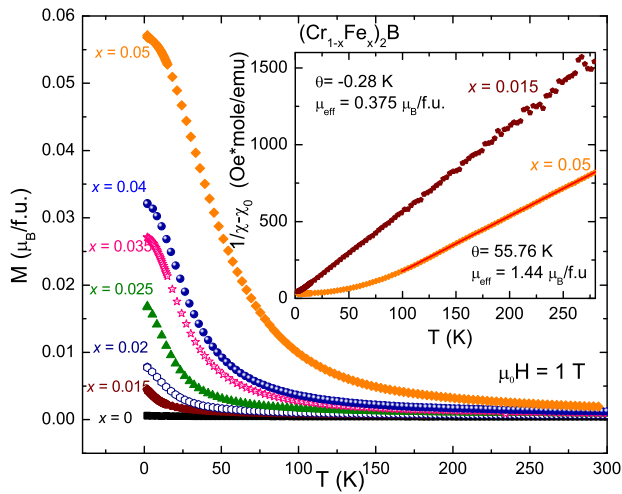}
  \caption{(color on line) Temperature dependent DC magnetization for (Cr$_{1-x}$Fe$_{x}$)$_2$B. Inset: the Curie-Weiss fits for representative samples. Data were taken in a standard zero-field cooled routine, under an applied magnetic field of $\mu_0 H = 1\,$T.}
\label{MT_CW}
\end{figure}

\begin{figure}[ht]
  \centering
  \includegraphics[width=\columnwidth]{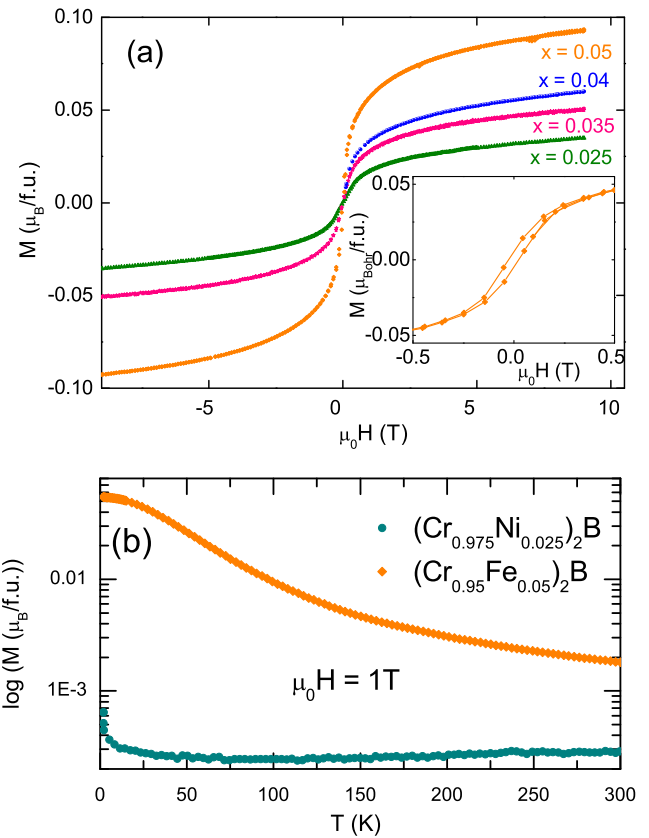}
  \caption{(color on line) Field dependent magnetization of (Cr$_{1-x}$Fe$_{x}$)$_2$B with higher doping. When \textit{x} = 0.05, a small hysteresis loop is observed (inset). (b) comparison of the effects of Fe and Ni doping. Note that both samples shown have the same valence electron count.}
\label{MH+Ni}
\end{figure}

No magnetic ordering was observed in pure Cr$_2$B above $2\,$K. Instead, Cr$_2$B displays temperature independent paramagnetism (with a small Curie tail at low temperature). Figure \ref{MT_CW} shows that if only 0.5 \% of the Cr is substituted by Fe, a small local moment appears. The local moment grows steadily with the amount of Fe substituted for Cr. Since metallic Fe and Fe$_2$B have Curie temperatures above room temperature, the observed moment cannot be due to impurities of these compounds. In contrast to Fe doping, Ni doping has no significant effect on the magnetism in Cr$_2$B (see Figure \ref{MH+Ni}(b)). (Cr$_{0.95}$Fe$_{0.05}$)$_2$B and (Cr$_{0.975}$Ni$_{0.025}$)$_2$B have the same valence electron count, but the latter does not exhibit any magnetic moment. This indicates that the magnetism in (Cr$_{1-x}$Fe$_{x}$)$_2$B cannot be attributed to a simple band filling effect.

%%%%%%%%%%%%%%%%%%%%%%%%%%%%%%%%%%%%%%%%%%%%%%%%%%%%%%%%%%%%%%%%%%%%%%
%%%%%%%%%%%%%%%%%%%%%%%%%%%%%%%%%%%%%%%%%%%%%%%%%%%%%%%%%%%%%%%%%%%%%%

Figure \ref{MH+Ni} shows the field dependent magnetization for samples with higher doping levels. The magnetization does not begin to saturate until fields near 9~T, a behavior that is usually associated with itinerant magnetism. At the highest doping level, (Cr$_{0.95}$Fe$_{0.05}$)$_2$B exhibits a small hysteresis. The coercive field is $22\,$mT, and the remnant magnetic moment is small, $0.005\,\mu_B$/f.u. 

%%%%%%%%%%%%%%%%%%%%%%%%%%%%%%%%%%%%%%%%%%%%%%%%%%%%%%%%%%%%%%%%%%%%%%
%%%%%%%%%%%%%%%%%%%%%%%%%%%%%%%%%%%%%%%%%%%%%%%%%%%%%%%%%%%%%%%%%%%%%%

\begin{figure}[ht]
  \centering
  \includegraphics[width=\columnwidth]{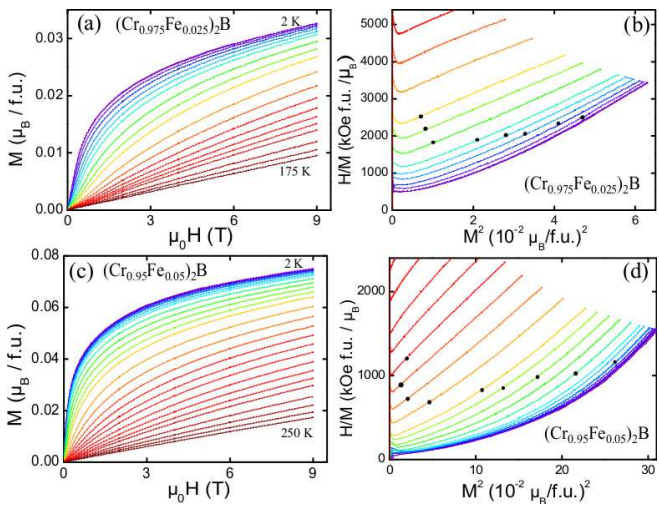}
  \caption{(color on line) Field depended magnetization at different temperatures (a) and Arrott plot (b) for (Cr$_{0.975}$Fe$_{0.025}$)$_2$B, and for (Cr$_{0.95}$Fe$_{0.05}$)$_2$B ((c) and (d)). The black dots indicate the field value above which the linear fits were performed to determine T$_C$.}
\label{MH_arrot}
\end{figure}

Figures \ref{MH_arrot} (a) and (c) show the field dependence of the bulk magnetization. In the paramagnetic high-temperature regime, characteristic linear M(H) is observed; at lower temperatures, however, significant curvature develops. In a mean field picture for an isotropic system, the magnetic contribution to the free energy to lowest order in the magnetization is $F = a/2 M^2 + b/4 M^4 +...$; it follows that $H = \partial F / \partial M = aM + bM^3$. Therefore, Arrott-plots of H/M as a function of M$^2$ allow us to explore the non-linear field dependence quantitatively. In agreement with this picture, the Arrott plots of our M(H) data are straight lines at high temperatures. Two possible scenarios may explain the significant low-field curvature of the Arrott plots at low temperatures. Firstly, it may be necessary to include the next-higher term in the equation of state, i.e. that $H/M = a+ bM^2 + gM^4$, which is particularly necessary when $b<0$ or is strongly temperature dependent \cite{smith2009magnetic}. This does not describe our observations well, however, particularly at \textit{x} = 0.05 and T=2~K. Secondly, a simple ferromagnetic picture may not be sufficient to describe the doping-induced magnetism in (Cr$_{1-x}$Fe$_{x}$)$_2$B; additional effects such as canting of the magnetic moments, which may reduce $M$  (and thus elevate $H/M$) from the value one might expect from a high-field extrapolation of the Arrott curves, may be present.

%%%%%%%%%%%%%%%%%%%%%%%%%%%%%%%%%%%%%%%%%%%%%%%%%%%%%%%%%%%%%%%%%%%%%%
%%%%%%%%%%%%%%%%%%%%%%%%%%%%%%%%%%%%%%%%%%%%%%%%%%%%%%%%%%%%%%%%%%%%%%

\begin{figure}[ht]
  \centering
  \includegraphics[width=\columnwidth]{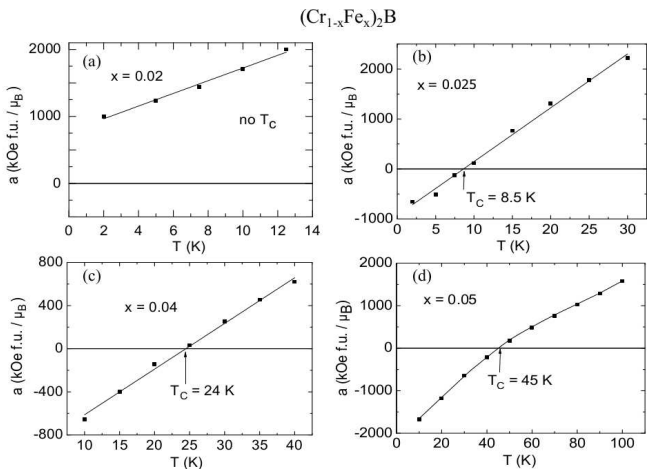}
  \caption{(color on line) Information inferred from the Arrott plots. The panels show the inverse susceptibilities $a = 1/\chi$ extracted from the Arrott plots, T$_C$ is the temperature where the inverse susceptibility intersects the abscissa.}
\label{arrottfits}
\end{figure}

\begin{figure}[ht]
  \centering
  \includegraphics[width=\columnwidth]{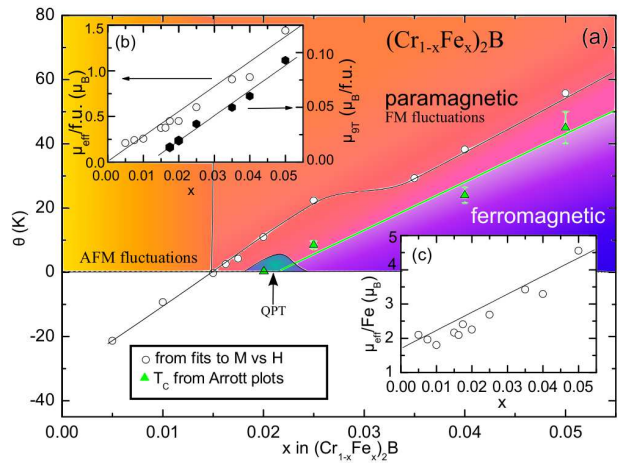}
  \caption{(color on line) Magnetic phase diagram for (Cr$_{1-x}$Fe$_{x}$)$_2$B. $\bigcirc$ symbols were obtained from fitting $M(T)$ data.  In panel (a), the green $\blacktriangle$ symbols show the T$_C$ obtained from the Arrott plot analysis (Figure \ref{arrottfits}). Panel (b) shows the values of the magnetization $\mu_{9T}$ at 2~K and 9~T (right axis). The left axis in panel (b) shows the effective magnetic moment per formula unit obtained from the Curie fit. In panel (c) this moment is shown per Fe atom rather than per formula unit.}
\label{pd}
\end{figure}

In order to extract a zero-field magnetic transition temperature T$_C$ as well as the inverse DC susceptibility $a = \chi^{-1}$ from the Arrott plots we focused on the high field data. (The black dots in Figures \ref{MH_arrot} (b) and (d) indicate the magnetic field value above which the fit was performed.) The result are shown in Figure \ref{arrottfits}, showing the inverse susceptibility (from the intercept of the fits) as a function of temperature for different \textit{x}. We define T$_C$ as the temperature where the inverse susceptibility $\chi^{-1} = a$ goes through 0. While no $T_C$ is found for \textit{x}=0.02, T$_C$ increases with doping, starting at \textit{x} = 0.025. To characterize the evolution of the fully polarized magnetic state as a function of doping \textit{x}, we define the quantity $\mu_{9T}$ as the magnetic moment at 2~K and 9~T.  $\mu_{9T}$ increases approximately linearly with doping between \textit{x} = 0.015 and 0.05 (Figure \ref{pd} (b)).

%%%%%%%%%%%%%%%%%%%%%%%%%%%%%%%%%%%%%%%%%%%%%%%%%%%%%%%%%%%%%%%%%%%%%%
%%%%%%%%%%%%%%%%%%%%%%%%%%%%%%%%%%%%%%%%%%%%%%%%%%%%%%%%%%%%%%%%%%%%%%

Figure \ref{pd} Panel (a) summarizes the information obtained from temperature and field dependent magnetization measurements in a phase diagram. Panel (b) shows the doping dependence of the fluctuating Curie-Weiss moment $\mu_{eff}$ per formula unit from Curie-Weiss fits to the data in Figure \ref{MT_CW}. The moment is small for \textit{x} = 0.005 (0.25 $\mu_B$) and increases nearly linearly with doping until a moment of 1.4 $\mu_B$ is reached for \textit{x} = 0.05. Panel (c) shows the effective moment per Fe atom, assuming that the complete moment resides on the Fe atoms. The moment for \textit{x} = 0.05 is larger than 4 $\mu_B$/Fe, which is very high and makes it unlikely that the magnetic moment originates from the Fe atoms alone; the Cr atoms are likely also being polarized. The main panel shows the doping dependency of the Weiss temperature $\Theta$. Cr$_2$B displays temperature independent paramagnetism. At the lowest Fe doping content, a negative Weiss temperature of - 20~K is observed, indicating rather strong antiferromagnetic interactions. The Weiss temperature increases with doping, and crosses 0~K at approximately \textit{x} = 0.015. (Cr$_{0.95}$Fe$_{0.05}$)$_2$B has a large positive Weiss temperature of about 50~K, indicating strong ferromagnetic interactions. Also shown in panel (a) are the Curie temperatures inferred from the Arrott plots. Panel (b) shows the doping dependence of $\mu_{9T}$, the moment at 2~K and 9~T. Again, a clear linear increase with doping is observed. $\mu_{9T}$ is more than an order of magnitude lower than the Curie-Weiss moments for all \textit{x}, where the ratio $\mu_{9T}/\mu_{eff} \sim 0.06$ is approximately independent of \textit{x}.

%%%%%%%%%%%%%%%%%%%%%%%%%%%%%%%%%%%%%%%%%%%%%%%%%%%%%%%%%%%%%%%%%%%%%%
%%%%%%%%%%%%%%%%%%%%%%%%%%%%%%%%%%%%%%%%%%%%%%%%%%%%%%%%%%%%%%%%%%%%%%

\subsection{Transport}
%%%%%%%%%%%%%%%%%%%%%%%%%%%%%%%%%%%%%%%%%%%%%%%%%%%%%%%%%%%%%%%%%%%%%%

All samples are metallic. Considering its polycrystalline nature, the residual resistance ratio $RRR = R(300\,$K$)/R(2\,$K$)= 27.3$ is relatively large for the stoichiometric sample. However, RRR already drops to $6.9$ when $0.5\,\%$ of the Cr atoms are replaced by Fe; and further to values just above 2 at higher dopant concentrations. This is most likely a direct consequence of the introduction of impurity scatterers. 

%%%%%%%%%%%%%%%%%%%%%%%%%%%%%%%%%%%%%%%%%%%%%%%%%%%%%%%%%%%%%%%%%%%%%%
\begin{figure}[ht]
  \centering
  \includegraphics[width=\columnwidth]{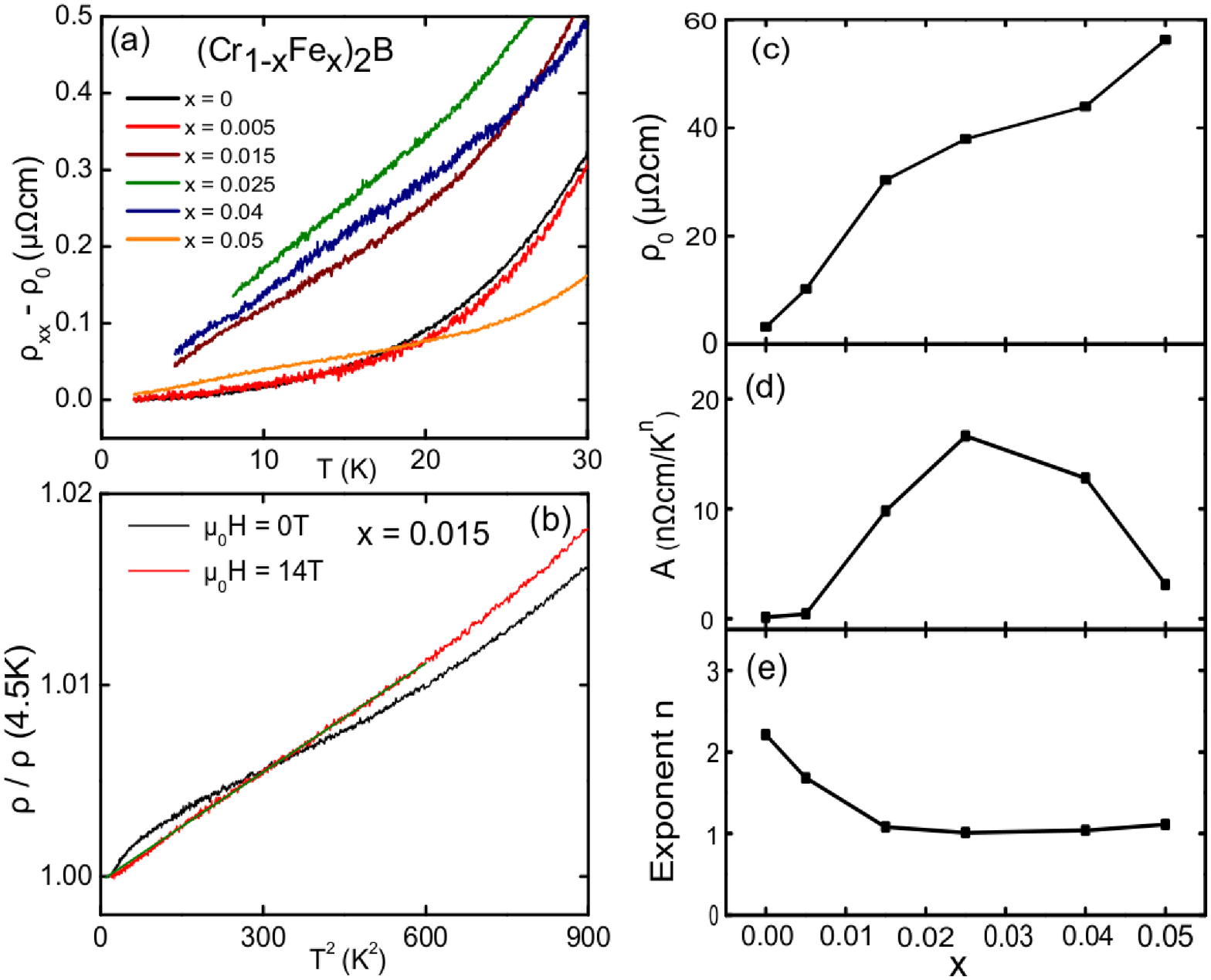}
  \caption{(color on line) (a) Low temperature electrical resistivity $\rho(T)$, corrected for $\rho_0$ as defined below. Quadratic Fermi liquid behavior is observed at small dopant concentrations, while linear behavior begins at \textit{x} = 0.015. For \textit{x} = 0.015, quadratic Fermi liquid behavior is recovered with a strong magnetic field $\mu_0 H = 14\,$T (b). Result of fitting the resistivity data to the standard power law expression $\rho = \rho_0 +AT^n$ (c-e). In (c), the residual resistance $\rho_0$ increases continuously as more impurities are introduced to the system. The $A$-parameter (d) peaks at \textit{x} = 0.025, an indication of enhanced electronic scattering around this concentration. Finally, in (e), the Fermi liquid $n=2$ exponent drops to a non-Fermi liquid value of $n \sim 1$.}
\label{transport}
\end{figure}
%%%%%%%%%%%%%%%%%%%%%%%%%%%%%%%%%%%%%%%%%%%%%%%%%%%%%%%%%%%%%%%%%%%%%%

The low-temperature behavior of $\rho(T)$ can be analyzed by a power-law of the form $\rho = \rho_0 + A T^n$. The Fermi-liquid exponent $n=2$ is common in metallic materials. In many materials, electronic or magnetic scattering dominates transport only below $2-10\,$K. However for stoichiometric Cr$_2$B and the lightly doped variants studied here, the Debye temperature is large ($\Theta_D \sim 800\,$K) and thus allows us to investigate the electronic transport unaffected by phonon scattering over an easily accessible temperature range. We show the results of the resistivity analysis in Figure \ref{transport} (right). In panel (c) of  figure \ref{transport}, $\rho_0$ is shown to increase continuously with dopant concentration, as is expected for impurity scattering. The values of $A$ obtained are shown in panel (d). The data show a distinct peak at intermediate dopant concentration, indicative of scattering from enhanced or critical fluctuations at the border of magnetism. At the same time, the exponent $n$, shown in panel (e), drops from the Fermi liquid value of 2 at low \textit{x} to an anomalous value of $n \sim 1$, where it remains at higher \textit{x}. 

The Fermi liquid exponent $n = 2$ is recovered by the application of a magnetic field of $\mu_0H = 14\,$T at \textit{x} = 0.015 (Figure \ref{transport} (b)); similar behavior is observed in the more strongly doped sample \textit{x} = 0.04, although much larger fields appear to be required to completely return to $n = 2$ in this case. In materials with enhanced magnetic fluctuations emanating from a low-temperature magnetic phase transition, non-Fermi liquid (nFL) $n=1$ behavior is typically observed in a fan-shaped section of the phase diagram, centered at the critical concentration $x_c$ \cite{gegenwart2008quantum}. In other words, Fermi liquid behavior at the lowest temperatures is recovered on both sides of $x_c$, with a cross-over to nFL behavior at a finite temperature $T_{FL}(x)$. In our measurements however, which are above 2K, the exponent in the resistivity remains $n=1$ up to the most strongly Fe-doped samples studied. We conclude that all our data was taken above $T_{FL}$ for the higher Fe concentrations and that measurements on samples with higher doping levels or transport experiments in the milli-Kelvin regime would be required to unambiguously confirm the recovery of the Fermi liquid exponent on the heavily doped side of the phase diagram.

\subsection{Heat Capacity}

The low temperature heat capacity of Cr$_2$B follows the expected C$_p$ = $\gamma$T + $\beta$T$^3$ relation, where $\gamma$T describes the electronic contribution to the heat capacity and $\beta$T$^3$ is the phonon contribution, which can be related to the Debye temperature $\Theta_D$ with the relation $\Theta_D = \left(\frac{12\pi^4nR}{5\beta}\right)^{1/3}$ with $n$ being the number of atoms per formula unit and $R$ being the ideal gas constant. $\gamma$ is the Sommerfeld parameter and is related to the density of states at the Fermi level normalized by an electron-phonon coupling parameter. For Cr$_2$B we find $\gamma$ = 11.11 mJ/(mole$\cdot$K$^2$ ) and $\Theta_D$ = 804~K (higher than the Debye temperature previously reported \cite{sirota2004thermodynamic}), for fitting the data in the range of 2-10~K. The lightly Fe-doped samples deviate from the C$_V$ = $\gamma$T + $\beta$T$^3$ relation at low T, as described below, but the Debye temperature from one composition to the next is not significantly different over such a small range of Fe doping. The Kadowaki-Woods ratio $A/\gamma^2 = 0.9\cdot 10^{-6} \mu \Omega $cm$($mol$\,$K$/$mJ$)^2$ of Cr$_2$B is consistent with the standard result for transition metal compounds \cite{Kadowaki1986507}.

%%%%%%%%%%%%%%%%%%%%%%%%%%%%%%%%%%%%%%%%%%%%%%%%%%%%%%%%%%%%%%%%%%%%%%
\begin{figure}[ht]
  \centering
  \includegraphics[width=\columnwidth]{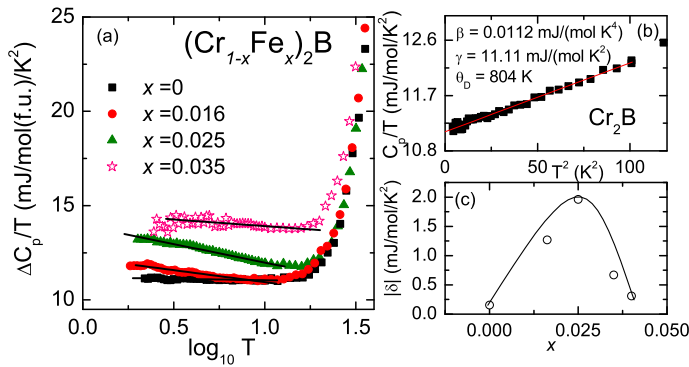}
  \caption{(color on line) $\Delta$ C/T plotted versus log T; a linear increase is observed for \textit{x} = 0.025. Panel (c) shows the dependence of the NFL prefactor $\delta$ which shows a maximum at \textit{x} = 0.025. Panel (b) displays the fit of the low T heat capacity for Cr$_2$B to C$_p$ = $\gamma$T + $\beta$T$^3$.}
\label{HC}
\end{figure}
%%%%%%%%%%%%%%%%%%%%%%%%%%%%%%%%%%%%%%%%%%%%%%%%%%%%%%%%%%%%%%%%%%%%%%

The Sommerfeld parameter $\gamma$ can be related to the Pauli paramagnetic susceptibility by: $\gamma = \frac{1}{3}\pi^2N_Ak_B^2D(E_F)$ and $\chi_p = \mu_B^2N_AD(E_F)$, where $N_A$ is Avogadro's constant, $k_B$ is Boltzmann's constant, $D(E_F)$ is the density of states at the Fermi level and $\mu_B$ is the Bohr magneton. This leads to the relation $\chi_p = 1.3715\cdot 10^{-5} \gamma$ where the unit of $\chi_p$ is emu/mol and the unit of $\gamma$ is mJ/(mole$\cdot$K$^2$). For Cr$_2$B this yields a Pauli paramagnetic susceptibility of $\chi_p = 1.5\cdot 10^{-4}$ emu/mol. Electronic structure calculations\cite{Cr2B_Magnetotransport} give a DOS of 3.1 states/eV/f.u. for Cr$_2$B, resulting in a (calculated) Pauli paramagnetic susceptibility of $\chi_p = 1.0\cdot 10^{-4}$ emu/mol - in good agreement with the value inferred from the heat capacity. The actual measured susceptibility for Cr$_2$B at 2~K however is $\chi_{2K} = 3.8\cdot 10^{-3}$ emu/mol, which is 25 times higher than the Pauli susceptibility inferred from both band structure calculations and the Sommerfeld coefficient. Hence Cr$_2$B is a strongly enhanced paramagnet. (Cr$_{0.95}$Fe$_{0.05}$)$_2$B has a susceptibility of $\chi_{max} = 3.0\cdot 10^{-2}$ emu/mol close to T$_C$, 200 times higher than $\chi_P$ determined from the heat capacity for Cr$_2$B. The resulting Stoner factor is strong evidence for the presence of a large thermodynamic weight of magnetic fluctuations \cite{brando2008logarithmic}.

Close to a ferromagnetic QCP the heat capacity is expected to follow $C_p/T = \delta ln(T) + \beta T^2$ ~\cite{PhysRevB.51.8996}. Figure \ref{HC} shows the electronic heat capacity plotted versus the logarithm of the Temperature for pristine and Fe-doped Cr$_2$B. The phonon contribution $\beta T^2$ obtained from the fit of pure Cr$_2$B was subtracted from the data. Linear behavior at low temperatures is only observed for \textit{x} = 0.025. Fitting the low temperature data reveals the magnitude of the prefactor $\delta$ of the nFL contribution. This prefactor clearly has a maximum at \textit{x} = 0.025 (left inset), which is in agreement with the resistivity data, where the \textit{x} = 0.025 sample has a maximum in $A$, and the magnetic data, which imply a paramagnetic to ferromagnetic critical point at the same composition. The absolute magnitude of the parameter $\delta$ for (Cr$_{1-x}$Fe$_x$)$_2$B ($\approx$ 2 mJ/(mole$\cdot$K$^2$ )) is comparable to the two other d-electron-based metallic weak ferromagnets where similar behavior has been observed, Pd$_{1-x}$Ni$_{x}$ \cite{nicklas1999non} and Nb$_{1+x}$Fe$_{2+x}$ \cite{brando2008logarithmic}; for these materials, the maximum value of $\delta$ is $\delta_{\text{max}} = 3.3$ and $\delta_{\text{max}} = 4.9$, respectively.

\subsection{Characterization of the Fe distribution}

It is natural to ask whether the observed behavior in this system can be attributed to the existence of small Fe clusters in the samples. In order to rule out this possibility, we employed state-of-the-art HRTEM and HAADF-STEM images and atomic resolution chemical analysis of (Cr$_{0.95}$Fe$_{0.05}$)$_2$B to map the Fe distribution in the sample.

%%%%%%%%%%%%%%%%%%%%%%%%%%%%%%%%%%%%%%%%%%%%%%%%%%%%%%%%%%%%%%%%%%%%%%
\begin{figure*}[ht]
  \centering
  \includegraphics[width=15cm]{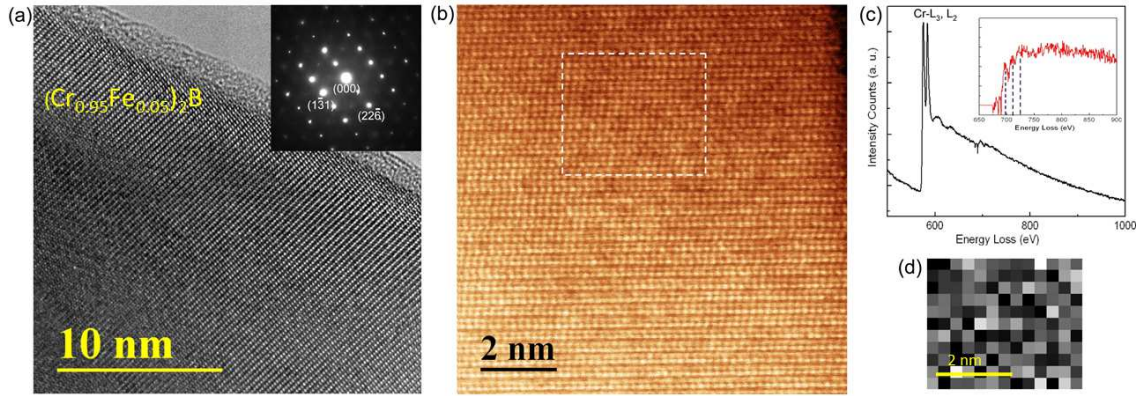}
  \caption{(color on line) HRTEM and HAADF-STEM images of (Cr$_{0.95}$Fe$_{0.05}$)$_2$B. (a) HRTEM image over a large length scale. No precipitates are seen, the inset shows the electron diffraction pattern of the same particle. (b) HAADF-STEM image of the same particle at higher magnification. No clusters are seen. (c) EELS results obtained from the same particle showing the Cr and Fe L-edges. The Cr L$_1$ edge, the Fe L$_3$ edge and L$_2$ edge, at 695 eV, 708 eV and 721 eV respectively, can be seen clearly in the inset with the background subtracted, positioned by the dash lines. (d) EELS mapping of the Fe distribution in the dash box in (b) where each pixel is roughly 3-4~\AA~ in size.  The brighter contrast corresponds to Fe atoms, showing that there is no clustering of Fe at the atomic level.}
\label{TEM}
\end{figure*}
%%%%%%%%%%%%%%%%%%%%%%%%%%%%%%%%%%%%%%%%%%%%%%%%%%%%%%%%%%%%%%%%%%%%%%

 Figure \ref{TEM} (a) shows a representative HRTEM image; the atoms are shown to be regularly ordered on the long range. No Fe-rich second phases are seen on this large length scale. The electron diffraction image (Figure \ref{TEM} (e)) is indexed to the [211] zone of the Cr$_2$B crystal structure. In (b) a HAADF-STEM image is shown at a higher magnification; also on this scale no clustering of Fe atoms is observed, and the atoms are very well ordered. Both (a) and (b) are typical of all images obtained for the sample; there are no Fe rich clusters present at the nanometer or larger scale. (Note: B is too light to be seen in STEM images therefore all atoms seen are Cr or Fe.) To look at the sample on the length scale of possible clusters consisting of only a few atoms, further analysis was performed, through atom-specific imaging enabled by EELS. Panel (c) shows a typical energy loss spectrum taken over an area at the nanoscale. Both Cr and Fe absorption edges are shown. The Fe intensities are weak, but visible. The Fe-L$_3$ and L$_2$ edges were then used to map the Fe in the sample on the scale of several atoms. The chemical mapping by EELS was performed in many areas over the sample and we present in Figure \ref{TEM} (d) a typical chemical analysis mapping obtained from an area about 4 x 5 nm$^2$ in size; the brighter contrast in the image represents individual Fe atoms. (Since the penetration depth of the scans is larger than just one atom different shades of gray can be seen in (d).) All such mappings showed a random Fe distribution. Thus the electron microscopy study gives very clear evidence that there is a random distribution of the Fe dopant in this sample. The magnetism therefore cannot originate from the presence of Fe clusters; the explanation for the observed behavior must involve the influence of individual Fe atoms on the surrounding Cr lattice.

%%%%%%%%%%%%%%%%%%%%%%%%%%%%%%%%%%%%%%%%%%%%%%%%%%%%%%%%%%%%%%%%%%%%%%
\section{Summary and conclusions}

We have reported the magnetic and transport properties of the intermetallic compound Cr$_2$B and its weakly Fe-doped analogues. We have shown that ferromagnetism can be induced with small amounts of Fe doping and that the doped Fe is randomly distributed. At the critical concentration x = 0.02, we observe linear behavior of the temperature dependence of the resistivity as well as what appears to be an anomalous logarithmic contribution to the heat capacity.

A logarithmic contribution to $C(T)$ is a hallmark signature of ferromagnetic quantum criticality in simple three-dimensional metals {\cite{Lonzarich1985} and has been reported for several lightly doped compounds such as Pd$_{1-x}$Ni$_x$ \cite{nicklas1999non} and Nb$_{1-x}$Fe$_{2+x}$ \cite{brando2008logarithmic}. Similarly, many strongly interacting (nearly magnetic) metals show a deviation from the standard Fermi liquid exponent $n=2$ in $\rho(T)$ \cite{gegenwart2008quantum}. Theoretical scenarios involving either purely ferromagnetic or coexisting ferro- and antiferromagnetic fluctuations have been brought forward to explain $n=1$ behavior in materials such as Ni$_3$Al \cite{niklowitz2005spin} and CePd$_2$Si$_2$ under pressure \cite{Mathur1998}. For Fe-doped Cr$_2$B, the change in sign of the Curie-Weiss temperature may indicate that antiferromagnetic and ferromagnetic fluctuations coexist over large portions of the phase diagram. 

Some of the signatures observed in our study may also be associated with the presence of dilute magnetic impurities in a non-magnetic host. For example, when small amounts of Fe impurities (x $<$ 3\%) are introduced into pure Au, linear resistivity and an enhancement of the heat capacity are seen \cite{macdonald1962thermo,harrison1967low}. Spin glasses may also display hysteretic magnetization isotherms. However, it is hard to understand the phenomenology observed here in the framework of spin-glass physics alone. Our Arrott analysis provides evidence for a low temperature magnetic phase transition as a function of doping, from the paramagnetic un-doped parent compound to bulk metallic magnetism incorporating both Fe-dopants and the Cr states. Moreover, the peaks of both $\delta$ and $A$ characterizing the transport properties and thermodynamic behavior close to the critical concentration indicate enhanced fluctuations close to the magnetic phase transition. The host of our experimental evidence points to a scenario not unlike the case of the strongly enhanced paramagnet Pd, where ferromagnetism and a QPT are induced by small amounts of Fe \cite{crangle1960ferromagnetism} or Ni \cite{nicklas1999non} dopants. 
%%%%%%%%%%%%%%%%%%%%%%%%%%%%%%%%%%%%%%%%%%%%%%%%%%%%%%%%%%%%%%%%%%%%%%%
\bigskip 
\begin{acknowledgments}

The materials preparation, structural characterization, and magnetic and specific heat measurements were supported by the DOE office of basic Energy Sciences, grant DOE FG02-08ER45706. The transport studies were supported by the NSF MRSEC program, grant DMR-0819860. The electron microscopy study was supported by the US Department of Energy (DOE)/Basic Energy Sciences, Materials Sciences and Engineering Division under Contract DE-AC02-98CH10886. The authors acknowledge helpful discussions with David Huse, F. Malte Grosche, Kai Sun and Neel Haldolaarachchige.  

\end{acknowledgments}

%%%%%%%%%%%%%%%%%%%%%%%%%%%%%%%%%%%%%%%%%%%%%%%%%%%%%%%%%%%%%%%%%%%%%%
\bibliography{Lit}

\end{document}